\documentclass[pra,twocolumn,superscriptaddress,amsmath]{revtex4}
\usepackage{orcidlink}

\newcommand{\ket}[1]{| #1 \rangle}
\newcommand{\braket}[2]{\langle #1 | #2 \rangle}
\newcommand{\ketbra}[2]{| #1 \rangle \langle #2 |}

\begin{document}
\title{Comment on ``Multiparty quantum mutual information: An alternative definition''}
\author{Jaehak Lee\,\orcidlink{0000-0002-3011-1129}}
\email{jaehak.lee.201@gmail.com}
\affiliation{Department of Physics and Astronomy, Seoul National University, Seoul 08826, Korea}
\affiliation{Center for Quantum Information, Korea Institute of Science and Technology, Seoul 02792, Korea}
\author{Gibeom Noh}
\affiliation{Department of Physics, Kyungpook National University, Daegu 41566, Korea}
\author{Changsuk Noh\,\orcidlink{0000-0001-8642-9104}}
\affiliation{Department of Physics, Kyungpook National University, Daegu 41566, Korea}
\author{Jiyong Park\,\orcidlink{0000-0002-3870-5714}}
\email{jiyong.park@hanbat.ac.kr}
\affiliation{School of Basic Sciences, Hanbat National University, Daejeon 34158, Korea}
\date{\today}

\begin{abstract}
We show that, contrary to the claim by Kumar [Phys. Rev. A \textbf{96}, 012332 (2017)], the quantum dual total correlation of an $n$-partite quantum state cannot be represented as the quantum relative entropy between $n-1$ copies of the quantum state and the product of $n$ different reduced quantum states for $n \geq 3$. Specifically, we argue that the latter fails to yield a finite value for generalized $n$-partite Greenberger-Horne-Zeilinger states.
\end{abstract}

\maketitle

In \cite{Kumar2017}, a quantum version of the dual total correlation \cite{Han1978} for an $n$-partite quantum state $\rho$ was proposed as
    \begin{equation} \label{eq:QDTC}
        I_{n} ( \rho ) \equiv \sum_{k = 1}^{n} S ( \rho_{\overline{k}} ) - ( n - 1 ) S ( \rho ),
    \end{equation}
where $S ( \tau ) = -\mathrm{tr} ( \tau \log \tau )$ is the von Neumann entropy of $\tau$ and $\rho_{\overline{k}} = \mathrm{tr}_{k} \rho$ denotes the $( n - 1 )$-partite quantum state obtained by taking the partial trace on the $k$th party of $\rho$. In addition, it was claimed that Eq. \eqref{eq:QDTC} can be represented as
    \begin{align} \label{eq:QDTCR}
        I_{n} ( \rho ) = J_{n} ( \rho ) \equiv S ( \rho^{\otimes ( n - 1 )} || \bigotimes_{k=1}^{n} \rho_{\overline{k}} ),
    \end{align}
where $\rho^{\otimes j}$ represents $j$ copies of $\rho$ and $S ( \tau || \sigma )$ is the quantum relative entropy of $\tau$ with respect to $\sigma$ \cite{Wilde2013},
    \begin{equation} \label{eq:RE}
        S ( \tau || \sigma ) = \begin{cases} \mathrm{tr} ( \tau \log \tau ) - \mathrm{tr} ( \tau \log \sigma ) & \mbox{if $\mathrm{supp} ( \tau ) \subseteq \mathrm{supp} ( \sigma )$} \\ \infty & \mbox{otherwise,} \end{cases}
    \end{equation}
where the support of $\omega$ is the Hilbert space spanned by the eigenstates of $\omega$ with non-zero eigenvalue \cite{NielsenChuang}.

It is well known that $I_{2} ( \rho )$ can be represented as the quantum relative entropy of a global quantum state $\rho$ with respect to the product of two local quantum states $\rho_{1}$ and $\rho_{2}$, i.e., $I_{2} ( \rho ) = S ( \rho || \rho_{1} \otimes \rho_{2} )$ \cite{Vedral2002}. We emphasize that the parties between global and local quantum states must be properly matched to avoid the infinity in Eq. \eqref{eq:RE}. While $\mathrm{supp} ( \rho ) \subseteq \mathrm{supp} ( \rho_{1} \otimes \rho_{2} )$ is always met, $\mathrm{supp} ( \rho ) \subseteq \mathrm{supp} ( \rho_{2} \otimes \rho_{1} )$ is not satisfied in general . It leads to a discrepancy between $I_{2} ( \rho ) = S ( \rho || \rho_{1} \otimes \rho_{2} ) $ and $J_{2} ( \rho ) = S ( \rho || \rho_{2} \otimes \rho_{1} )$. For instance, we have $I_{2} ( \rho ) = 0$ and $J_{2} ( \rho ) = \infty$ for $\rho = \ketbra{\psi_{1}}{\psi_{1}} \otimes \ketbra{\psi_{2}}{\psi_{2}}$ satisfying $\braket{\psi_{1}}{\psi_{2}} = 0$.

One can generalize this observation to the case of $n \geq 3$. Looking into the order of the parties in $\rho^{\otimes (n-1)}$ and $\bigotimes_{k=1}^{n} \rho_{\overline{k}}$, one immediately sees that they are mismatched for all $n$ because the former and the latter start with the first and second parties of $\rho$, respectively. Similar to the case of $n=2$, if we look into a product state $\rho = \bigotimes_{k=1}^{n} \ketbra{\psi_{k}}{\psi_{k}} $ satisfying $\braket{\psi_{i}}{\psi_{j}} = \delta_{i,j}$, we obtain $I_{n} ( \rho ) = 0$ and $J_{n} ( \rho ) = \infty$.

One may think that rearranging the parties of quantum states can resolve the support condition problem. It is possible for $n=2$ but impossible for $n \geq 3$. We show this by investigating the case of $\rho = \ketbra{\phi}{\phi}$ with $\ket{\phi} = \sqrt{p} \ket{0}^{\otimes n} + \sqrt{1-p} \ket{1}^{\otimes n}$. The state of $\ket{\phi}^{\otimes (n-1)}$ is represented by the superposition of the basis states having the multiples of $n$, i.e., $\{ 0, n, 2n, ..., n(n-1) \}$, copies of $\ket{1}$. On the other hand, the eigenstates of $\bigotimes_{k=1}^{n} \rho_{\overline{k}}$ have multiples of $n-1$, i.e., $\{ 0, (n-1), 2(n-1), ..., n(n-1) \}$, copies of $\ket{1}$. As $n$ and $n-1$ are coprimes, $\ket{\phi}^{\otimes (n-1)}$ is orthogonal to the eigenstates of $\bigotimes_{k=1}^{n} \rho_{\overline{k}}$ except for $\ket{0}^{\otimes n(n-1)}$ and $\ket{1}^{\otimes n(n-1)}$, which means $\mathrm{supp} ( \rho^{\otimes (n-1)} ) \not\subseteq \mathrm{supp} ( \bigotimes_{k=1}^{n} \rho_{\overline{k}} )$ for $n \geq 3$. Importantly, the observation just described remains valid even if we rearrange the parties. Therefore, we have $I_{n} ( \rho ) = - n [ p \log p + (1-p) \log (1-p) ]$ and $J_{n} ( \rho ) = \infty$ independent of the matching structure between $\rho^{\otimes (n-1)}$ and $\bigotimes_{k=1}^{n} \rho_{\overline{k}}$.
 
In \cite{Kumar2017}, the equivalence between Eqs. \eqref{eq:QDTC} and \eqref{eq:QDTCR}, i.e., $I_{n} ( \rho ) = J_{n} ( \rho )$, was invoked in order to prove the non-negativity and monotonicity of  $I_{n} ( \rho )$ under a partial trace and completely positive maps.  In addition, the proof for non-negativity was reproduced in \cite{Guo2023}. Our counterexample invalidates these proofs which rely on the facts that the relative entropy is non-negative and non-increasing under completely positive and trace-preserving (CPTP) maps. Interestingly, however, one can find alternative proofs for the non-negativity and monotonicity in \cite{Cerf2002}. In Ref. \cite{Cerf2002}, $I_{n} ( \rho )$ was proposed as a quantum secrecy monotone, and was shown to be non-negative and non-increasing under local CPTP maps, owing to strong subadditivity of the conditional quantum mutual entropy. Therefore, $I_{n} ( \rho )$ is a suitable monotonic measure of multi-partite correlations, while $J_{n} ( \rho )$ is not.

\clearpage
J.L. and J.P. acknowledge support from the National Research Foundation of Korea funded by the government of Korea (MSIT) (Grants No. NRF-2022M3K4A1097117 and No. NRF-2019R1G1A1002337, respectively).

\bibliographystyle{apsrev}

\end{document}